\documentclass{actasen}
\newcommand{\Slash}[1]{{\ooalign{\hfil/\hfil\crcr$#1$}}}
\usepackage{epstopdf}
\begin{document}

\setcounter{page}{115}

\Volume{2011}{35}


\runheading{Ryo Yoshiike, Kazuya Nishiyama, Toshitaka Tatsumi}%

\title{Spontaneous Magnetization of Quark Matter in Inhomogeneous Chiral Phase}






\enauthor{Ryo Yoshiike, Kazuya Nishiyama, Toshitaka Tatsumi}{Department of Physics, Kyoto Univeresity, Kyoto 606-8502, Japan}

\abstract{
Considering the density wave of scalar and pseudoscalar condensates, we study the response of quark matter to a weak external magnetic field. In an external magnetic field, the energy spectrum of the lowest Landau level becomes asymmetric about zero, which is closely related to chiral anomaly. This spectral asymmetry gives rise to spontaneous magnetization. This mechanism may be one of candidates for the origin of the strong magnetic field in magnetars. Furthermore, using the generalized Ginzburg-Landau(gGL) expansion, we show that magnetic susceptibility exhibits a peculiar feature.}


\maketitle

\section{Introduction}

Recently, the existence of the inhomogeneous chiral phase in QCD phase diagram has been suggested by the analysis of chiral effective model such as the NJL model \cite{nakano, nickel} or Schwinger-Dyson approach \cite{muller}. In this phase, the quark condensates periodically modulate. Here, we consider ``dual chiral density wave"(DCDW) \cite{nakano}, where the quark condensates take the form,
\begin{align}
 \Delta(\bm{r}) \equiv \langle \bar{\psi} \psi \rangle + i \langle \bar{\psi} i\gamma^5 \tau_3 \psi \rangle = \Delta e^{iqz}, \label{condensate}
\end{align}
within the two-flavor case. 
In the DCDW phase, quarks exhibit an interesting feature in the presence of the magnetic field through spectral asymmetry \cite{tatsumi}. In the following, we analyse the response of the quark matter in DCDW phase to the magnetic field and show the possibility that the quark matter has the spontaneous magnetization, which may be related to the strong magnetic field in magnetars \cite{skokov,olausen}. \par
The origin of the strong magnetic field in compact stars, especially in magnetars, has been one of the long-standing problems. As the candidates of the origin, amplification of the magnetic field by dynamo mechanism, magnetorotational instability or the hypothesis of the fossil magnetic field has been proposed from the macroscopic point of view. However, no definite conclusions have been obtained. The possibility of spontaneous magnetizatin in the quark matter gives another mechanism from the microscopic point of view.

\section{MODEL and FORMALISM}

We use the two-flavor NJL model in the mean field approximation,
\begin{align}
 {\cal{L}}_{\rm MF} =& \bar\psi \left\{i\Slash{D} +2G\left[\langle \bar{\psi} \psi \rangle + i\gamma^5\tau_3 \langle \bar{\psi} i\gamma^5 \tau_3 \psi \rangle\right]\right\} \psi + G\bigl[\langle\bar\psi\psi\rangle^2 + \langle\bar\psi i\gamma^5\tau_3\psi\rangle^2\bigr]. \label{njl}
\end{align}
Here, the chiral limit is taken and we take the external magnetic field $\bm{B}$ along the $z$ axis. Mean fields take the DCDW condensate (\ref{condensate}). Then, the energy spectrum constitutes the Landau levels and
the lowest Landau level(LLL) becomes asymmetric about zero \cite{frolov}. According to \cite{niemi}, spectral asymmetry generates anomalous particle number, $N_{{\rm anom}}= -\frac{1}{2}\sum_k {\rm sign}(\lambda_k)$, where $\lambda_k$ is the energy spectrum of LLL. Note that, this anomalous particle number in the DCDW phase \cite{tatsumi} agrees with the contribution of chiral anomaly by the Wess-Zumino-Witten term \cite{son}. To investigate the response of the quark matter to the weak magnetic field, thermodynamic potential at zero temperature is expanded about the magnetic field $B$, $\Omega(\mu,B\,;m,q) = \Omega^{(0)}(\mu\,;m,q)+eB\Omega^{(1)}(\mu\,;m,q)+\cdots$, where $\mu$ represents the chemical potetial.
Remarkably, $\Omega^{(1)}$ emerges only when energy spectrum has asymmetry and includes the contribution of anomaly \cite{tatsumi}.

\section{SPONTANEOUS MAGNETIZATION} 
\begin{figure}[h]
 \centering
 \begin{tabular}{c}
      \begin{minipage}{0.5\hsize}
        \begin{center}
          \includegraphics[width=4cm, angle=270]{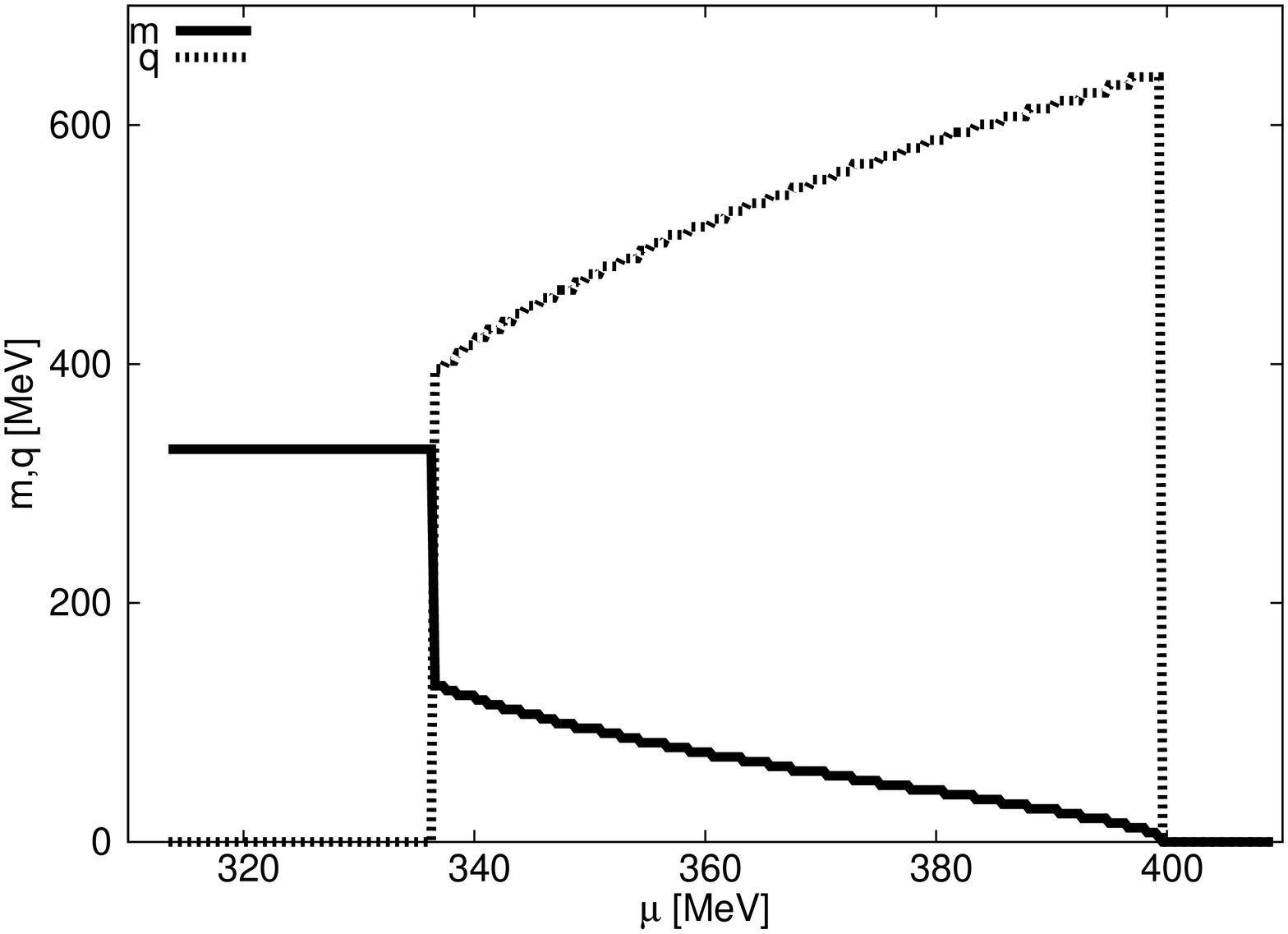}
          \hspace{1.6cm} (a) 
        \end{center}
      \end{minipage}

      \begin{minipage}{0.5\hsize}
        \begin{center}
          \includegraphics[width=4cm, angle=270]{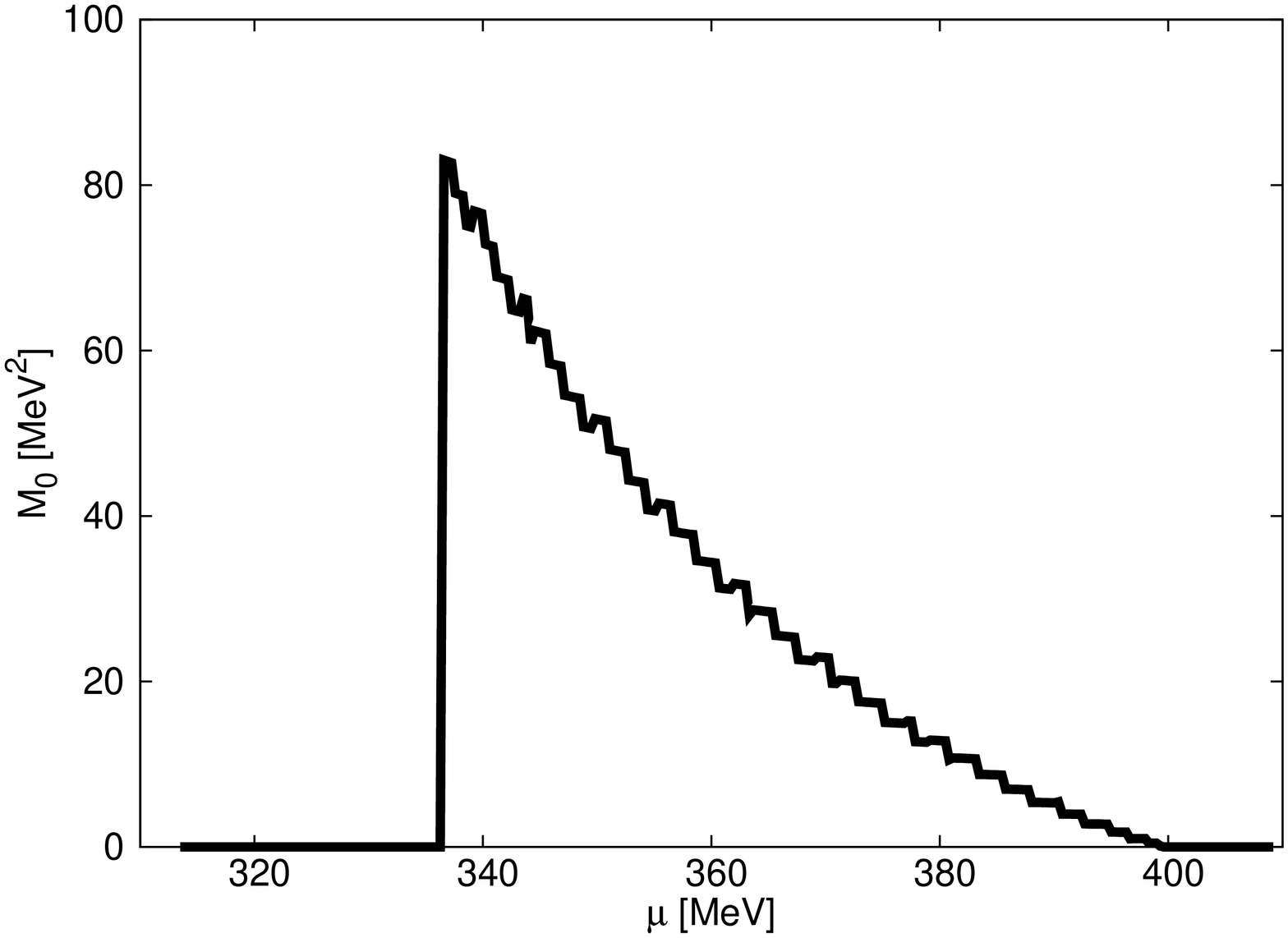}
          \hspace{1.6cm} (b)
        \end{center}
      \end{minipage}      
 \end{tabular}
 \caption{Chemical potential dependence of (a) the order parameters at $B=0$ and (b) spontaneous magnetization.}
 \label{magnetization}
\end{figure}

Order parameters, $m$ and $q$, are determined by minimizing thermodynamic potential. Using the minimized thermodynamic potential $\Omega_{{\rm min}}(\mu,B)$, spontaneous magnetization is given by,
$ M_0 \equiv -\partial \Omega_{\rm min}/\partial B\big|_{B=0} = -e\Omega^{(1)}\big(\mu\,;m=m^{(0)}(\mu),q=q^{(0)}(\mu)\big).$
Here, $m^{(0)}$ and $q^{(0)}$ minimize the thermodynamic potential at $B=0$. Numerical result is represented in Fig. \ref{magnetization}. The region with $m\neq 0,q=0$ represents the homogeneously broken phase, that with $m\neq 0, q\neq 0$ represents the DCDW phase and that with $m=0$ represents the chirally restored phase. As one can see in Fig. \ref{magnetization}, spontaneous magnetization becomes nonezero only in the DCDW phase. Assuming the sphere of quark matter with DCDW, the magnetic field produced by the spontaneous magnetiztion is roughly estimated as $10^{16}{\rm G}$ on the surface. 

\section{magnetic susceptibility}
Generally, thermodynamic potential can be expanded about order parameters around the critical point. Using the gGL expansion for spatially dependent condensates \cite{ggl}, we can see that the term proportional to $\bm{B}$ in the DCDW phase \cite{tatsumi}. In this case, the critical point becomes the Lifshitz point(LP) where the inhomogeneous chiral phase, the homogeneously broken phase and the chirally restored phase intersect. \par
\begin{figure}[h]
 \centering
 \includegraphics[width=5cm, angle=270]{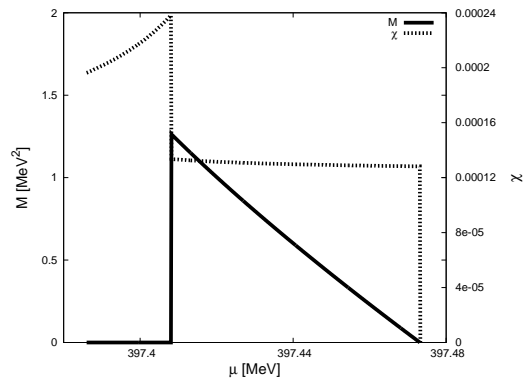}
 \caption{Chemical potential dependence of spontaneous magnetization($M$) and magnetic susceptibility($\chi$) at $T\simeq 100{\rm MeV}$}
 \label{susceptibility}
\end{figure}

Magnetization is obtained by the same way of section 3 and magnetic susceptibility is given by $\chi \equiv \frac{\partial M}{\partial B}\big|_{B=0} = -\frac{\partial^2 \Omega_{\rm min}}{\partial B^2}\big|_{B=0}$. Numerical results at $T\simeq 100{\rm MeV}$, very close to LP, are drawn in Fig. \ref{susceptibility}. As one can see in Fig. \ref{susceptibility}, the magnetic susceptibility does not diverge but is discontinuous at the phase transition points. This is a peculiar feature in the DCDW case, different from the usual ferromagnetism.    

\acknowledgements{The authors thank T.-G. Lee for useful discussions.}


\end{document}